\documentclass[conference]{IEEEtran}
\IEEEoverridecommandlockouts
\usepackage{cite}
\usepackage{amsmath,amssymb,amsfonts}
\usepackage{algorithmic}
\usepackage{rotating}
\usepackage{lscape}
\usepackage{graphicx}
\usepackage{textcomp}
\usepackage{xcolor}
\usepackage{multirow}
\usepackage{adjustbox}
\usepackage{float}
\usepackage{subcaption}

\def\BibTeX{{\rm B\kern-.05em{\sc i\kern-.025em b}\kern-.08em
    T\kern-.1667em\lower.7ex\hbox{E}\kern-.125emX}}
    
\begin{document}

\title{Quantifying Influencer Impact on Affective Polarization}

% \author{\IEEEauthorblockN{Anonymous Author}
% \vspace{0.8in}
% }

\author{\IEEEauthorblockN{Rezaur Rashid}
\IEEEauthorblockA{\textit{Department of Computer Science} \\
\textit{UNC Charlotte}\\
Charlotte, USA \\
mrashid1@charlotte.edu}
\and

\IEEEauthorblockN{Joshua Melton}
\IEEEauthorblockA{\textit{Department of Computer Science} \\
\textit{UNC Charlotte}\\
Charlotte, USA \\
jmelton30@charlotte.edu}
\and

\IEEEauthorblockN{Ouldouz Ghorbani}
\IEEEauthorblockA{\textit{Department of Computer Science} \\
\textit{UNC Charlotte}\\
Charlotte, USA \\
oghorban@charlotte.edu}
\and

\IEEEauthorblockN{Siddharth Krishnan}
\IEEEauthorblockA{\textit{Department of Computer Science} \\
\textit{UNC Charlotte}\\
Charlotte, USA \\
skrishnan@charlotte.edu}
\and

\IEEEauthorblockN{Shannon Reid}
\IEEEauthorblockA{\textit{Department of Criminal Justice and Criminology } \\
% \IEEEauthorblockA{\textit{\color{red}Dept. of Criminal Justice and Criminology } \\
\textit{UNC Charlotte}\\
Charlotte, USA \\
sreid33@charlotte.edu}
\and

\IEEEauthorblockN{Gabriel Terejanu}
\IEEEauthorblockA{\textit{Department of Computer Science} \\
\textit{UNC Charlotte}\\
Charlotte, USA \\
gabriel.terejanu@charlotte.edu}}

\maketitle

\begin{abstract}
% In an era where digital platforms increasingly mediate public discourse, grasping the complexities and nuances in affective polarization—especially as influenced by key figures on social media—has never been more vital. This study delves into the intricate web of interactions on Twitter, now rebranded as 'X', to unravel how influencer-led conversations catalyze shifts in public sentiment, laying bare the complex dynamics that underpin online polarization. Employing a novel methodological framework that includes counterfactual analysis, we analyze scenarios with and without specific influencer-led conversations. Our findings illuminate the significant role influencers play in shaping public discourse, offering insights into the mechanisms of online polarization and suggesting pathways for future research to mitigate divisiveness and explore new methods for quantifying affective polarization. This research contributes to the broader understanding of digital communication's impact on societal polarization, underscoring the importance of detailed analysis in developing strategies to foster a more cohesive digital public sphere.

In today’s digital age, social media platforms play a crucial role in shaping public opinion. This study explores how discussions led by influencers on Twitter, now known as ‘X’, affect public sentiment and contribute to online polarization. We developed a counterfactual framework to analyze the polarization scores of conversations in scenarios both with and without the presence of an influential figure. Two case studies, centered on the polarizing issues of climate change and gun control, were examined. Our research highlights the significant impact these figures have on public discourse, providing valuable insights into how online discussions can influence societal divisions. 

%In an era where digital platforms increasingly mediate public discourse, grasping the complexities and nuances in affective polarization—especially as influenced by key figures on social media—has never been more vital. This study delves into the intricate web of interactions on Twitter, now rebranded as 'X', to unravel how influencer-led conversations catalyze shifts in public sentiment, laying bare the complex dynamics that underpin online polarization. In this work, we proposed a framework that analyzes conversations of influential figures in specific scenarios both with and without a conversation. Our findings illuminate the significant role influencers play in shaping public discourse, offering insights into the mechanisms of online polarization and suggesting pathways for future research to mitigate divisiveness and explore new methods for quantifying affective polarization. This research contributes to the broader understanding of digital communication's impact on societal polarization, underscoring the importance of detailed analysis in developing strategies to foster a more cohesive digital public sphere.
\end{abstract}


% INTRO
\section{Introduction}
% The transformative impact of social media on political communication cannot be overstated. As a platform for instantaneous information exchange, social media has redefined the way individuals engage with political content, debate policy, and form community bonds~\cite{dahlgren2009media, chambers2013social}. However, this digital revolution has also given rise to a less auspicious phenomenon: affective polarization~\cite{yair2020note, yu2021affective, rudolph2021affective}. This term encapsulates the growing emotional divide between individuals with opposing political stances, characterized not only by differences in policy opinion but also by deep-seated animosity and distrust~\cite{feldman2023affective, serrano2021digital}.
The transformative impact of social media on political communication cannot be overstated. As a platform for instantaneous information exchange, social media has redefined the way individuals engage with political content, debate policy, and form community bonds~\cite{dahlgren2009media, chambers2013social}. However, this digital revolution has also given rise to a less auspicious phenomenon: affective polarization~\cite{yair2020note, yu2021affective, rudolph2021affective}. The term, affective polarization refers to the phenomenon where individuals feel more positively towards members of their political group while simultaneously harboring negative sentiments towards those of opposing groups~\cite{feldman2023affective, serrano2021digital}.

Affective polarization on social media is especially harmful because it can undermine healthy democratic discussions. Instead of encouraging thoughtful debate, it often leads to hostile confrontations~\cite{harel2020normalization, jenke2023affective}. The effects of this trend go beyond the online world, influencing real-world political engagement and the broader societal fabric~\cite{tyagi2020affective, beel2022linguistic}.

%Affective polarization on social media is particularly pernicious due to its potential to erode the foundations of democratic discourse, replacing reasoned debate with hostile confrontation. It transforms disagreement into contempt, making compromise and consensus-building increasingly elusive~\cite{harel2020normalization, jenke2023affective}. The implications of this trend extend beyond the digital realm, influencing real-world political engagement and the broader societal fabric~\cite{tyagi2020affective, beel2022linguistic}.

At the heart of this polarization are influential users—individuals and entities with large followings who can significantly shape public discourse. These influencers can either deepen existing divides or help bridge gaps in understanding through their discussions on controversial topics~\cite{johnson2020issues, balietti2021reducing}. Nowadays, the reach and impact of such figures are greater than ever before, making it essential to closely examine their role in the process of affective polarization.~\cite{recuero2019using}.

In this study, we aim to measure the impact of conversations started by influential users on the polarization seen in Twitter/X discussions. Our paper makes the following contributions to understand how these influencers shape public sentiment and contribute to polarization in online communities.

%In this study, we aim to quantify the effects of conversations initiated by influential users on the polarization observed in Twitter discourses. Leveraging computational methods, we dissect the emotional tone and polarization of social media conversations, tracing the shifts in sentiment that occur when influencer-led conversations are introduced or removed from the discourse. By doing so, we seek to provide a granular understanding of how influencers sway public sentiment and contribute to the emotional polarization of online communities.

%Our paper makes the following contributions:

\begin{itemize}
    \item This research presents a new framework using counterfactual analysis to quantitatively measure how conversations led by influencers affect polarization on Twitter/X. By comparing polarization scores with and without these influential conversations, we reveal their impact on public discourse.
    
    %\item This research introduces a framework utilizing the concept of counterfactual analysis to quantitatively assess how conversations led by influencers affect polarization on Twitter/X. 
    
    %\item We implement a novel methodology that compares polarization scores with and without the presence of influential conversations, shedding light on the direct impact of these discourses.

    \item We offer a detailed analysis of how influential users shape emotional dynamics within contentious topics such as climate change and gun control, providing quantifiable measures of their influence on affective polarization.

\end{itemize}

This study not only advances our understanding of affective polarization but also has practical implications for the design and governance of social media platforms. By identifying the influential figures and the factors that exacerbate or mitigate polarization, platform designers and policymakers can better navigate the challenges posed by this phenomenon and work towards a more informed and less divided public discourse.

% LIT REVIEW
\section{Literature Review}

%%% Refined Lit Rev.
The phenomenon of affective polarization has been extensively documented in political psychology, with recent studies revealing its escalation on social media platforms~\cite{tornberg2021modeling, gillani2018me}. Affective polarization extends beyond ideological disagreements, encapsulating emotional responses that manifest as mutual dislike and distrust among those with opposing political allegiances~\cite{druckman2022mis}.

% The literature on this subject reveals a complex ecosystem where media, political figures, and ideological silos interact to deepen divisions. Social media platforms, with their algorithmically curated content, have been implicated as amplifiers of this division, fostering environments where inflammatory rhetoric and moral outrage thrive~\cite{suarez2022toxic, heatherly2017filtering, mason2016cross}. Echo chambers, a topic of much debate within the literature, are said to contribute to polarization by reinforcing ideological conformity and limiting exposure to diverse viewpoints~\cite{cinelli2021echo}. While some studies suggest that these echo chambers might insulate users from contrary opinions, thus exacerbating polarization, others argue against this insularity, suggesting that exposure to opposing views does not necessarily diminish polarization and may, in some cases, intensify it~\cite{lee2022social, iyengar2019origins}.

Research highlights a complex interplay between media, political figures, and entrenched ideologies that magnifies societal divisions. Social media platforms enhance these divisions through algorithmically curated content that often promotes inflammatory and polarizing material~\cite{suarez2022toxic, heatherly2017filtering, mason2016cross}. Echo chambers, a topic of much debate, are often criticized for reinforcing ideological conformity and shielding users from opposing viewpoints, which can intensify affective polarization~\cite{cinelli2021echo}. Conversely, some research suggests that even when individuals are exposed to contrary opinions, this exposure does not necessarily mitigate polarization and may, under certain conditions, actually exacerbate it~\cite{lee2022social, iyengar2019origins}.

Advancements in quantifying affective polarization have led to the development of metrics that capture the emotional content and toxicity in social media interactions~\cite{tyagi2020affective, tyagi2021heated}. Studies now classify users by political partisanship and analyze the emotions and language used in their communications, providing a subtle understanding of the affective component of polarization~\cite{feldman2023affective, connors2023social}. This line of research highlights that negative emotions and toxicity are not randomly distributed but correlate with the network distance in social media interactions, suggesting structural properties of these networks influence the emotional tone of online discourse~\cite{nordbrandt2023affective}.

Kramer et al.\cite{kramer2014experimental} assert that emotional states are contagious so whether expressed by other people or appears on Newsfeed (a personalized stream of content provided by social media platforms) has a direct impact on our emotions. Another finding of their study is that as opposed to prevalent perception, non-verbal language, and inter-personal interactions are not required for contagion of emotions. The study by Cha et al.\cite{cha2010measuring} explores influence patterns on social media and identifies the most important role-playing factors. 
Betts and Bliuc~\cite{betts2022effect}  shows that an influencer with extreme opinions will invariably accelerate the pace of polarization, and this impact grows in proportion to their influence and level of engagement. The impact of a neutral influencer, however, depends on the society's openness to differing viewpoints.

Despite extensive research on affective polarization, much of the current literature focuses on the consequences of polarization without a clear methodology to quantify its emergence and escalation directly from influencer interactions. In addition, there is a limited exploration into the specific role of influential users within these divisive dynamics, particularly in how their conversations shape public sentiment over time~\cite{recuero2019using}. 

Our research builds upon these findings by specifically examining how influential figures impact the affective landscape of online discourses. By systematically evaluating the presence and absence of high-profile conversations, we offer a unique perspective on the role influencers play in either mitigating or exacerbating affective polarization. Through this lens, we contribute to a deeper understanding of the dynamics at play in social media's political discourse to provide insights into digital communication strategies aimed at reducing polarization.

% METHODOLOGY
\section{Methodology}

\begin{figure*}
    \centering
    \includegraphics[width=1.80\columnwidth]{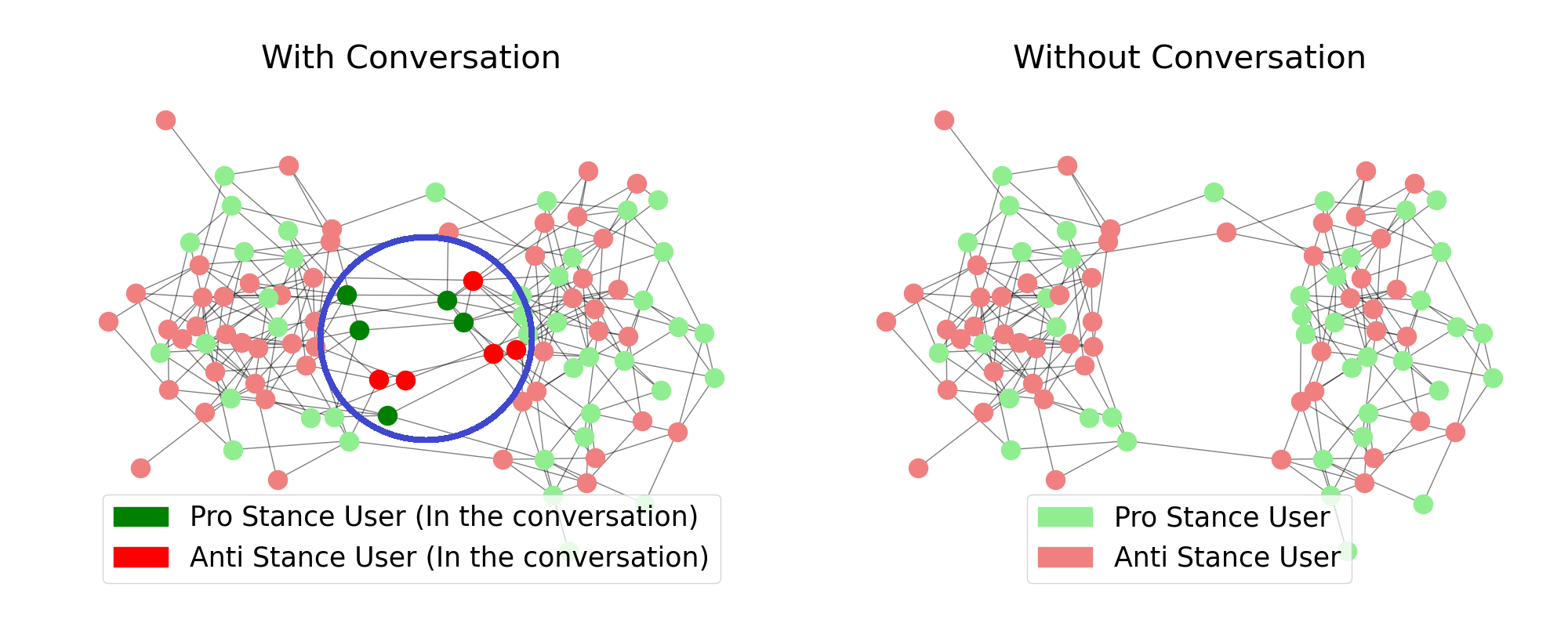}
    % \caption{Illustrative network diagram of an interaction network with and without a conversation,  highlighting the structural differences. The left panel includes the conversation (circled), whereas the right panel excludes this conversation. The dense areas indicate ongoing interactions through other conversations among the influencer's followers.}
    \caption{Interaction networks with and without a specific conversation. The left panel displays the network including the conversation (circled), while the right panel shows the network without it, highlighting the dense areas where ongoing interactions among the influencer's followers persist through other conversations.}
    \label{fig:example_graph}
\end{figure*}

The goal of this study is to methodically quantify the impact of influencers on affective polarization on social media platforms. To achieve this, we implement a counterfactual analysis framework, which involves constructing hypothetical scenarios to understand what might happen if certain influencer-led conversations did not occur. This approach, illustrated in Figure~\ref{fig:example_graph}, allows us to isolate the effects of these conversations on polarization dynamics, providing a clearer picture of their influence. The following subsections introduce the specifics of our data collection strategies, the construction of interaction networks, and the application of sentiment analysis and quantifying polarization metrics, providing a comprehensive understanding of our methodological approach and the analytical techniques employed to evaluate the influencer impact on online social interactions.

\subsection{Data}

\subsubsection{Data Collection}

% Our research utilized a comprehensive dataset extracted from Twitter, encompassing a wide array of tweets related to two prominent and contentious political issues--climate change and gun control. We used Hashtags to scrape the initial set of relevant tweets related to each topic. To fully capture conversation cascades and user interactions, we then recursively collect all referenced tweets from each tweet in the initially matched set. This ensures that all available tweets from relevant conversations are included in our analysis, even if particular tweets do not explicitly use any of the hashtags. We also retrieved user metadata, including the number of likes, replies, retweets, etc. to identify influential users, as shown in Table~\ref{table-dataset-details}.

Our research employed a comprehensive dataset, we extracted from Twitter, covering tweets related to two prominent and contentious political issues--climate change and gun control. We used specific Hashtags relevant to each topic to scrape the initial set of tweets using the Twitter API before its restrictions were imposed in 2023. To ensure a thorough capture of conversation cascades and user interactions, we expanded our data collection by recursively retrieving all referenced tweets linked to the initial set. This method allowed us to include all pertinent discussions, capturing the depth and breadth of conversations. User metadata such as the number of likes, replies, and retweets was also retrieved, as shown in Table~\ref{table-dataset-details}.

%%%%% Table Datasets
\begin{table*}[h!]
\small
\centering
\caption{Dataset details}
\begin{adjustbox}{width=\textwidth}
% \usepackage{multirow}

\begin{tabular}{|l|l|r|r|r|r|r|r|}
\hline
\textbf{Dataset} & \textbf{Timeline} & \begin{tabular}[c]{@{}r@{}} \textbf{Total Tweets} \\ \textbf{(million)}\end{tabular} & \begin{tabular}[c]{@{}r@{}}\textbf{Total Conversations}\\ \textbf{(million)}\end{tabular} & \begin{tabular}[c]{@{}r@{}}\textbf{Total Users}\\ \textbf{(million)}\end{tabular} & \begin{tabular}[c]{@{}r@{}}\textbf{\#likes}\\ \textbf{(billion)}\end{tabular} & \begin{tabular}[c]{@{}r@{}}\textbf{\#replies}\\ \textbf{(million)}\end{tabular} & \begin{tabular}[c]{@{}r@{}}\textbf{\#retweets}\\ \textbf{(billion)}\end{tabular} \\ \hline
Gun Control & 2022.01.01 - 2022.12.31 & 2.65  & 2.26  & 0.94 & 1.46 & 120.69  & 7.49 \\ \hline
Climate Change & 2021.06.01 - 2022.05.31 & 7.24  & 6.46  & 2.04 & 30.39 & 149.05  & 2.64 \\ \hline
\end{tabular}

\end{adjustbox}
\label{table-dataset-details}
\end{table*}

% %%% TABLE Influential users
% \begin{table*}[h!]
% \small
% \centering
% \caption{Top 5 Influential Users for each dataset}
% \begin{adjustbox}{width=1.75\columnwidth}
% \begin{tabular}{|l|l|r|r|r|r|l|}
% \hline
% \textbf{Dataset} & \textbf{Twitter Account} & \textbf{Tweet Count} & \begin{tabular}[c]{@{}r@{}}\textbf{Total Likes}\\ \textbf{(million)}\end{tabular} & \begin{tabular}[c]{@{}r@{}}\textbf{Total Retweets}\\ \textbf{(million)}\end{tabular} & \begin{tabular}[c]{@{}r@{}}\textbf{Total Replies}\\ \textbf{(million)}\end{tabular} & \textbf{Stance} \\ \hline
% \multirow{5}{*}{Gun Control} & User 1 & 1,241 & 272.40  & 25.55  & 19.69  & anti \\ \cline{2-7} 
%  & User 2 & 1,665 & 66.79  & 10.35  & 10.77  & pro \\ \cline{2-7} 
%  & User 3 & 1,750 & 37.74  & 4.75  & 2.75  & pro \\ \cline{2-7} 
%  & User 4 & 1,760 & 31.75  & 5.07  & 4.15  & pro \\ \cline{2-7} 
%  & User 5 & 746 & 22.81  & 5.25  & 1.15  & pro \\ \hline
% \multirow{5}{*}{Climate Change} & User 1 & 1,766 & 190.47  & 16.64  & 11.84  & believe \\ \cline{2-7} 
%  & User 2 & 76 & 185.99  & 48.82  & 9.99  & believe \\ \cline{2-7} 
%  & User 3 & 1,945 & 57.79  & 8.05  & 6.79  & believe \\ \cline{2-7} 
%  & User 4 & 33 & 26.43  & 7.53  & 0.65  & believe \\ \cline{2-7} 
%  & User 5 & 55 & 24.12  & 5.61  & 1.14  & believe \\ \hline
% \end{tabular}

% \end{adjustbox}
% \label{table-influential users}
% \end{table*}

%%% TABLE Influential users
\begin{table}[h!]
\small
\centering
\caption{Top 5 influential users for each dataset}
% \begin{adjustbox}{width=0.95\columnwidth}
\begin{tabular}{|l|l|r|r|l|}
\hline
\textbf{Dataset} & \textbf{User} & \begin{tabular}[c]{@{}r@{}}\textbf{Tweet} \\ \textbf{Count}\end{tabular} & \begin{tabular}[c]{@{}r@{}}\textbf{Total Likes}\\ \textbf{(million)}\end{tabular} & \textbf{Stance} \\ \hline
\multirow{5}{*}{\begin{tabular}[c]{@{}l@{}}Gun-\\ Control\end{tabular}} & User 1 & 1,241 & 272.40   & anti \\ \cline{2-5} 
 & User 2 & 1,665 & 66.79    & pro \\ \cline{2-5} 
 & User 3 & 1,750 & 37.74    & pro \\ \cline{2-5} 
 & User 4 & 1,760 & 31.75    & pro \\ \cline{2-5} 
 & User 5 & 746 & 22.81    & pro \\ \hline
\multirow{5}{*}{\begin{tabular}[c]{@{}l@{}}Climate-\\ Change\end{tabular}} & User 1 & 1,766 & 190.47    & believe \\ \cline{2-5} 
 & User 2 & 76 & 185.99    & believe \\ \cline{2-5} 
 & User 3 & 1,945 & 57.79    & believe \\ \cline{2-5} 
 & User 4 & 33 & 26.43    & believe \\ \cline{2-5} 
 & User 5 & 55 & 24.12    & believe \\ \hline
\end{tabular}

% \end{adjustbox}
\label{table-influential users}
\end{table}

\subsubsection{Data Exploration}

 % A preliminary assessment revealed that despite the high volume of conversations captured within our datasets,  a notable portion of these conversations are limited in engagement as depicted in Figure~\ref{fig:tweet_distributions}. This observation indicates that a substantial quantity of conversations remains solitary, devoid of further interaction that could potentially influence the polarization of social discourse. In light of this, our analysis is confined to threads manifesting a minimum engagement criteria of 20 tweets and 10 distinct users.

\begin{figure}[ht]
    \begin{subfigure}{1.1\columnwidth}
        \centering
        \includegraphics[width=1\columnwidth]{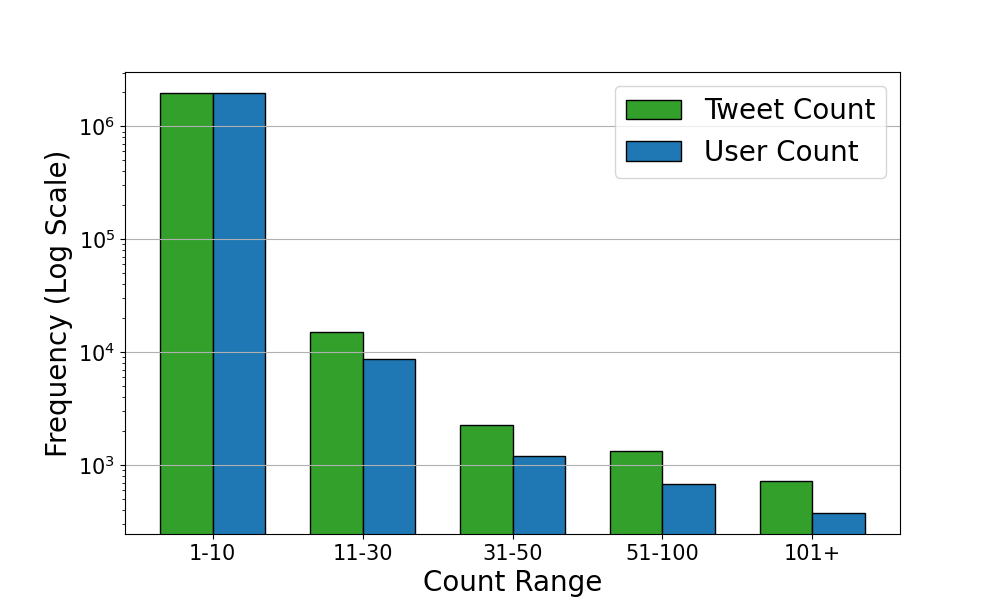}
        \caption{Gun Control Dataset}
        \label{fig:gun_tweets}
    \end{subfigure}

    \begin{subfigure}{1.1\columnwidth}
        \centering
        \includegraphics[width=1\columnwidth]{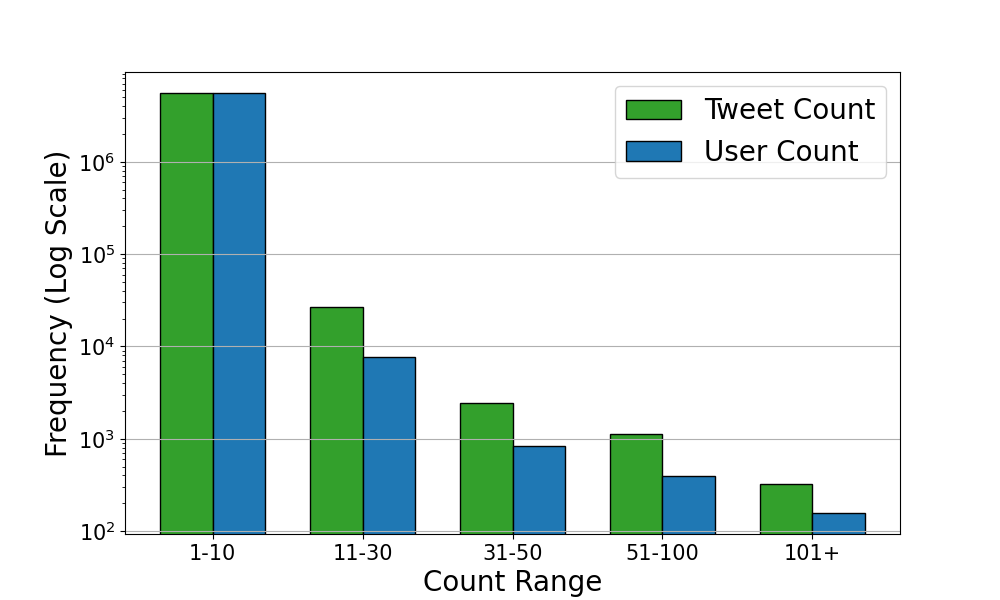}
        \caption{Climate Change Dataset}
        \label{fig:climate_tweets}
    \end{subfigure}
    
    % \caption{Distribution of numbers of tweets per conversation and numbers of users per conversation: (a) Gun Control and (b) Climate Change Topics.}
    \caption{Conversation characteristics: (a) Gun Control and (b) Climate Change topics, displaying the frequency of conversations by number of tweets (left) and number of users (right).}
    
    \label{fig:tweet_distributions}
\end{figure}

 In analyzing the dataset, we identified a significant number of conversation threads with limited active engagement, defined as the number of distinct users replying within each thread. This measure focuses on direct interactions rather than passive activities like retweets and likes, aiming to capture meaningful exchanges. Many threads exhibited minimal interaction, often involving less than ten participants  (Figure~\ref{fig:tweet_distributions}). To ensure the analytical robustness of our study, we established minimum engagement criteria based on discussions with domain experts. Only threads with at least 20 tweets and participation from at least 10 distinct users were included.

\subsubsection{Influential Users}
% The study incorporates a selection of the top influential users from each dataset, identified by their capacity to attract substantial engagement, as evidenced by metrics such as accumulated likes (e.g. Table~\ref{table-influential users} presents a representative subset). This methodology is predicated on the premise that these opinion leaders wield a considerable impact on the discourse permeating online platforms. Despite the extensive user base encapsulated within our datasets, the chosen influencers serve as a microcosm for wider interaction trends, with their content demonstrating a significant resonance amongst a broader audience. From these users adhering to our engagement criteria, we honed in on 5,500 gun control and 8,421 climate change conversations for a detailed examination which resulted in 2,103 influential users for the gun control dataset and 3,701 for the climate change dataset. This analysis of influencer-initiated conversations over an annual span enables us to discern the dynamics of influencer-driven polarization and to extrapolate these findings to reflect overarching tendencies on the platform.

Our methodology involved identifying influential users within each dataset based on their ability to engage substantial audience interactions, primarily measured by the total number of likes their posts received. Influential users were determined by sorting all users in descending order based on their like counts and selecting those at the top of this list (Table~\ref{table-influential users}), which indicates a significant impact on online discourse. Although Table~\ref{table-influential users} shows only the top five for illustrative purposes, our analysis included a broader set of influential figures, with 2,103 users for the gun control dataset and 3,701 for the climate change dataset. We examined 5,500 gun control and 8,421 climate change conversations initiated by these users over a period of one year.

%, we gain insights into the resonant themes and interaction patterns that potentially guide the broader trends of online polarization.

\subsection{Quantifying Polarization with E/I Index}

Our study employs a multi-model sentiment analysis approach using VADER, BERTweet, and RoBERTa to evaluate the emotional content of tweets. This comprehensive sentiment profiling forms the basis for our subsequent analysis of affective polarization.

We utilize the E/I Index methodology, adopted from Tyagi et al. \cite{tyagi2020affective}, to quantify affective polarization by measuring the ratio of external (between-group) to internal (within-group) interactions within Twitter discourse networks. This method allows us to assess the degree of in-group cohesion and out-group engagement. The E/I Index is derived by evaluating the balance of positive and negative interactions within and between groups. We compute the difference between the E/I indices of positive and negative interactions to discern the polarization valence. For further details on the computations and equations underlying the E/I Index, we refer readers to the original work by Tyagi et al.\cite{tyagi2020affective}.

This approach is akin to the P-Index proposed by Guerra et al.~\cite{guerra2013measure}, albeit tailored specifically for Twitter’s unique interaction dynamics. Through this framework, we focus on:
\begin{itemize}
\item \textbf{In-group Solidarity:} Gauged through the frequency and sentiment of interactions within a group.
\item \textbf{Out-group Engagement:} Characterized by the nature of interactions across different stance groups, often highlighting conflict or disagreement.
\end{itemize}

In practical applications, such as in gun control debates, this index helps us identify and describe patterns like Pro-Anti and Anti-Pro, indicating shifts from passive support or opposition to active advocacy or contention. While Bestvater et al.~\cite{bestvater2023sentiment} caution against using sentiment as a standalone measure for polarization or stance, our study integrates these insights as part of a broader, multi-faceted approach to understanding affective polarization.

\subsection{User Stance Labeling using Graph Neural Networks (GNNs)}
Our approach to user stance labeling employs a two-stage pipeline that integrates textual and social interaction data, as established in~\cite{melton2024two}.

% \paragraph{Initial Label Generation}
% \label{section: initial label gen}
% Initially, we deploy a reciprocal label propagation algorithm within a user-hashtag bipartite graph. This method effectively generates labels by associating users with probable stance categories based on their interactions with specific hashtags. This labeling serves as a flexible precursor to more definitive stance categorization, accommodating the preliminary subtleties and complexities of user stances. Notably, this approach yields stance labels for a limited subset of users, compared to a significantly larger population, and does not take into account any social interactions between users.

% \paragraph{Expanding Stance Labeling with GNNs}

% We expand stance labeling by constructing a comprehensive user-user interaction network, where nodes represent users and edges signify connections between authors and targets of their interactions. We make use of BERTweet to embed textual content into node features. We then train a graph neural network classifier on this interaction graph to predict node labels, using the seed users labeled by the hashtag method as the training set. This approach is grounded in the principles of linguistic and network homophily~\cite{yang2017overcoming, mcpherson2001birds, kipf2016semi}, which provide a theoretical foundation for understanding how users' language and interactions reflect their ideological alignments. Moreover, the GNN predicts node labels by generating a probability distribution over possible stances for each user.

\paragraph{Initial Label Generation}
\label{section: initial label gen}
Our stance labeling process begins with the construction of a user-hashtag bipartite graph, where one set of nodes represents users and the other set represents hashtags used in their tweets. This graph forms the basis for applying a reciprocal label propagation algorithm, initially assigning stance labels based on users' engagements with specific hashtags. We identify seed labels from a subset of users who frequently use hashtags that are strongly associated with known stances. This preliminary method generates labels for approximately 500,000 users in the gun control dataset and 1.6 million users in the climate change dataset, providing an initial but incomplete picture of user stances, and does not take into account any social interactions between users.

\paragraph{Expanding Stance Labeling with GNNs}

To enhance and refine our stance labeling, we construct a comprehensive user-user interaction network. In this network, nodes represent users and edges signify direct interactions such as retweets or replies. Leveraging BERTweet, we embed the textual content of tweets into node features to capture linguistic nuances. Subsequently, we employ a Graph Neural Network (GNN) classifier (GraphSAGE), trained on the interaction graph with seed labels derived from the initial hashtag-based method. This two-stage approach integrates both textual content and interaction patterns to predict user stances more accurately, resulting in expanded stance labels for 2.6 million users in the gun control context and 4.7 million users in the climate change context. Moreover, the GNN predicts node labels by generating a probability distribution over possible stances for each user. Detailed information on the GNN's architecture, parameter settings, and performance evaluation can be found in the work by Melton et al.~\cite{melton2024two}.

\paragraph{Optimization of Classification Thresholds}

% To find the users in a stance group, we employed a dual-threshold approach, utilizing Threshold\_1 and Threshold\_2 to categorize users based on their stance probability. Through a systematic grid search, we optimized the threshold combination to minimize misclassification near the decision boundary, maximizing the F1 score. This involved comparing predicted labels against known heuristic stances from the hashtag method and a subset of manually labeled users with domain expertise. The optimal thresholds were established as Threshold\_1 ($\le0.40$) for 'pro' or 'believers' stances and Threshold\_2 ($\ge0.60$) for 'anti' or 'disbelievers', with intermediate probabilities classified as 'undecided'. Consequently, a significant number of users and their initiated conversations were excluded from both datasets due to their stance being labeled as 'undecided'.

To more accurately delineate users into distinct stance groups, we adopted a dual-threshold strategy for classifying user stances based on their stance probability scores. We identified optimal thresholds, Threshold\_1 and Threshold\_2, through a comprehensive grid search methodology designed to refine the decision boundary by optimizing the F1 score. This process entailed meticulously comparing predicted labels against a benchmark set comprising both heuristic stances derived from our initial hashtag-based labeling and a subset of users manually annotated by domain experts. This dual-source validation approach ensured robustness in threshold selection by integrating empirical data with expert judgment. The final thresholds were determined as follows: Threshold\_1 ($\leq 0.40$) for identifying 'pro' or 'believers' stances, and Threshold\_2 ($\geq 0.60$) for 'anti' or 'disbelievers'. Users with probabilities falling between these thresholds were classified as 'undecided', and subsequently, their data were excluded from further analysis in both datasets.

\subsection{Approach to Quantifying Influencer Impact on Affective Polarization}

%Our study employs a multifaceted approach to quantify the influence of key Twitter influencers on affective polarization, integrating comparative analyses with strategic subgraph constructions. 

We center our methodology around the concept of a counterfactual scenario within a subgraph of influence, where polarization scores are calculated in scenarios both with and without specific influencer-led conversations, as illustrated in Figure~\ref{fig:example_graph}. This comparative method is crucial for isolating the shifts in polarization attributable to these conversation networks, providing insights into how individual conversations can sway public sentiment.

To enhance our analysis, we focus on constructing subgraphs around specific influencers and their follower networks, recognizing that the comprehensive interaction graphs on platforms like Twitter, which involve millions of daily interactions, can obscure the effects of smaller-scale conversation networks. For instance, in our detailed case study of 'User X's Twitter interactions, we selectively identified a subset of approximately 1,000 followers from a larger pool of 15,000 active users. We constructed a subgraph that included these followers and their adjacent connections, encompassing around 2,000 users in total.

This subgraph methodology provides a focused lens through which we can observe and analyze the polarization dynamics more clearly. Initially, we compute the polarization within this subgraph by including 'User X's conversation. Subsequently, to distinctly understand the conversation's specific impact, we recalibrated the polarization after removing the 400 users directly involved in the conversation. This technique of contrasting polarization scores with and without the conversation offers a transparent view of how specific conversations influence the network's polarization dynamics.

By focusing on influencer-centric subgraphs, we uncover the subtle yet significant impacts of specific conversations on affective polarization, avoiding the dilution of findings by the broader network's noise.

% By focusing on influencer-centric subgraphs, we can uncover the subtle yet significant impacts of specific conversations on affective polarization. This approach ensures that our findings are not diluted by the noise of the broader network, providing a clearer understanding of the pivotal role these influencers play in either mitigating or exacerbating affective polarization. Through this refined analytic lens, we contribute to a deeper understanding of the dynamics at play in social media's political discourse.

% EXPERIMENTAL RESULTS
\section{Results}

\begin{figure}[ht]
    \begin{subfigure}{1\columnwidth}
        \centering
        \includegraphics[width=1\linewidth]{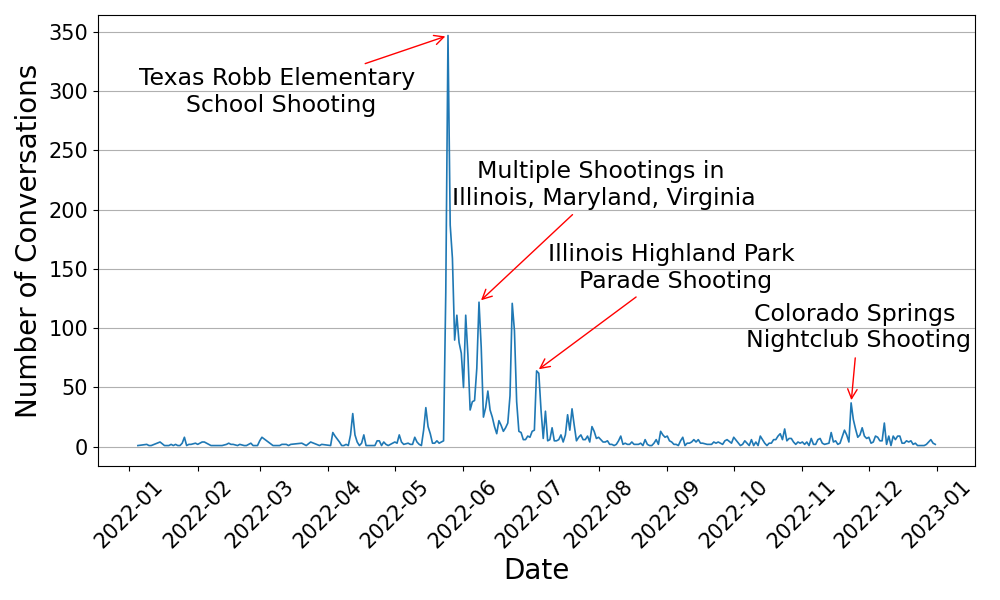}
        \caption{Gun Control Dataset}
        \label{fig:gun_tweets: conv}
    \end{subfigure}

    \begin{subfigure}{1\columnwidth}
        \centering
        \includegraphics[width=1\linewidth]{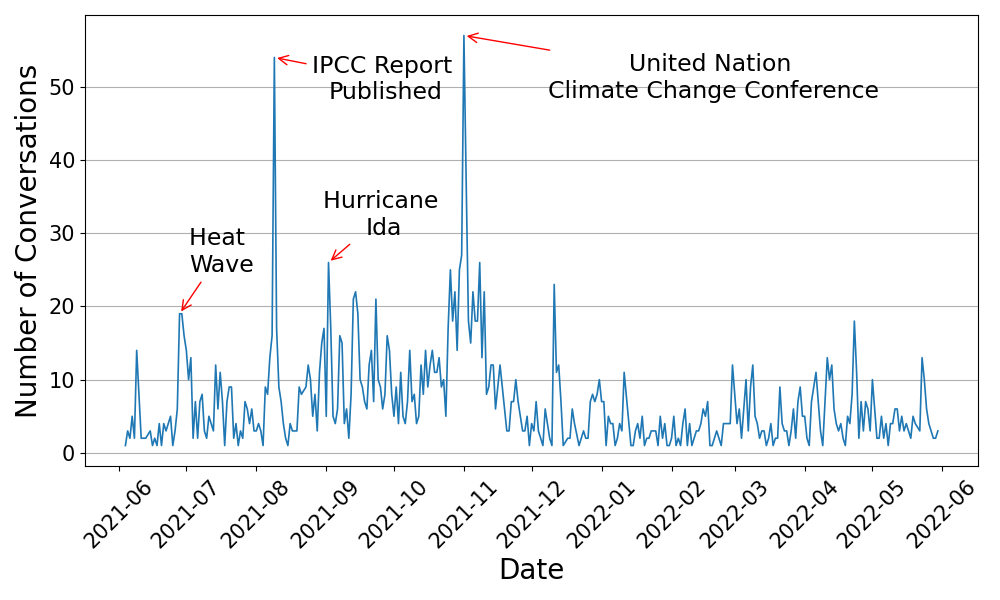}
        \caption{Climate Change Dataset}
        \label{fig:climate_tweets: conv}
    \end{subfigure}

    \caption{Daily conversation counts over the specified time frame, stratified by topic: (a) Gun Control and (b) Climate Change.}
    % \caption{Distribution of the number of conversations per day over the specified period: (a) Gun Control and (b) Climate Change Topics.}
    \label{fig:conversation_distribution}
\end{figure}

\begin{figure*}[ht]
    % First row of subfigures
    \begin{subfigure}[b]{0.32\textwidth}
        \centering
        \includegraphics[width=\linewidth]{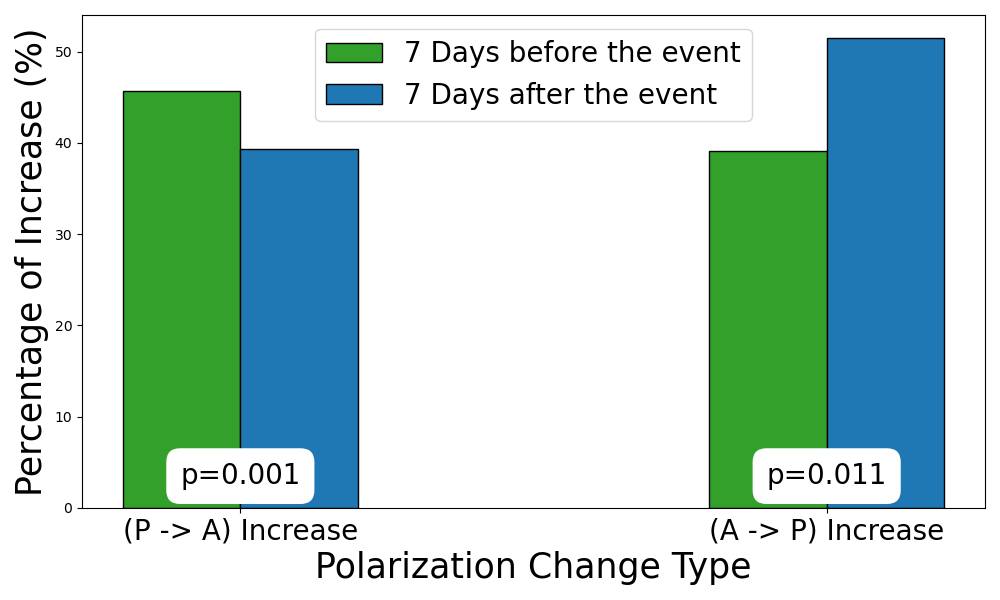}
        \caption{Texas Shooting}
        \label{fig:plot1}
    \end{subfigure}
    \hfill % Fills space between the subfigures
    \begin{subfigure}[b]{0.32\textwidth}
        \centering
        \includegraphics[width=\linewidth]{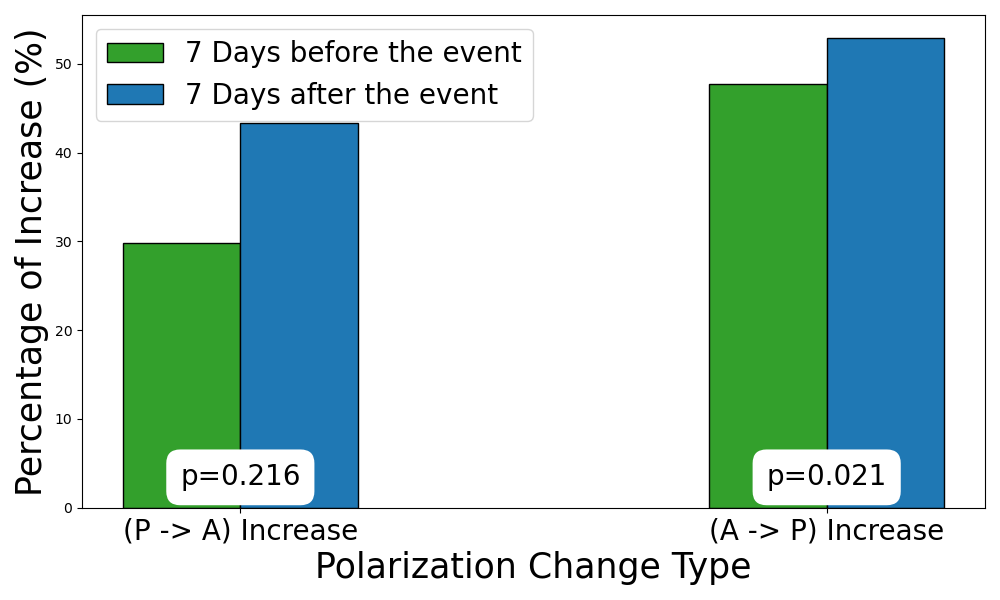}
        \caption{Illinois Shooting}
        \label{fig:plot2}
    \end{subfigure}
    \hfill
    \begin{subfigure}[b]{0.32\textwidth}
        \centering
        \includegraphics[width=\linewidth]{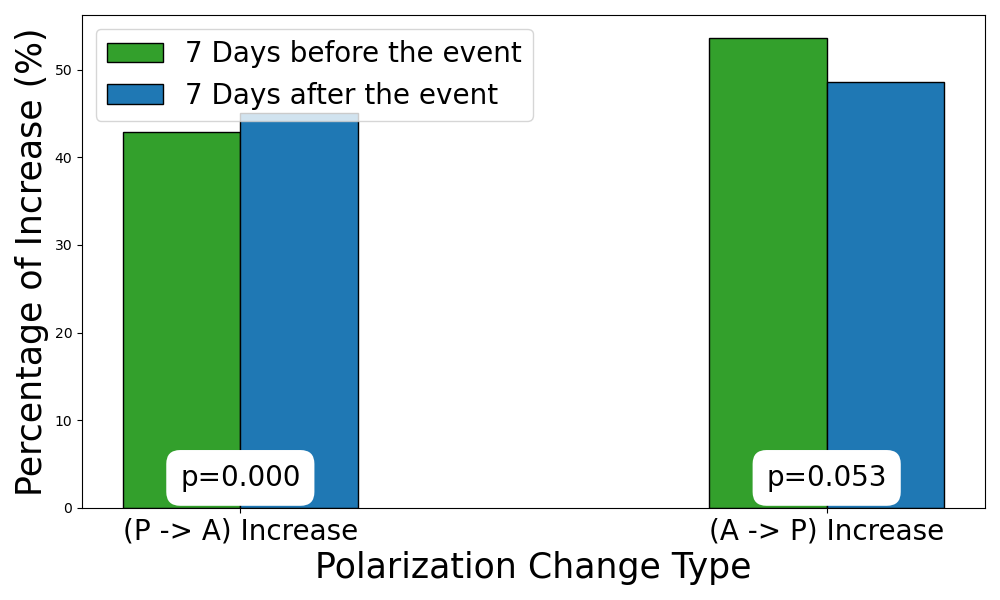}
        \caption{Colorado Shooting}
        \label{fig:plot3}
    \end{subfigure}
    
    % Second row of subfigures
    \begin{subfigure}[b]{0.32\textwidth}
        \centering
        \includegraphics[width=\linewidth]{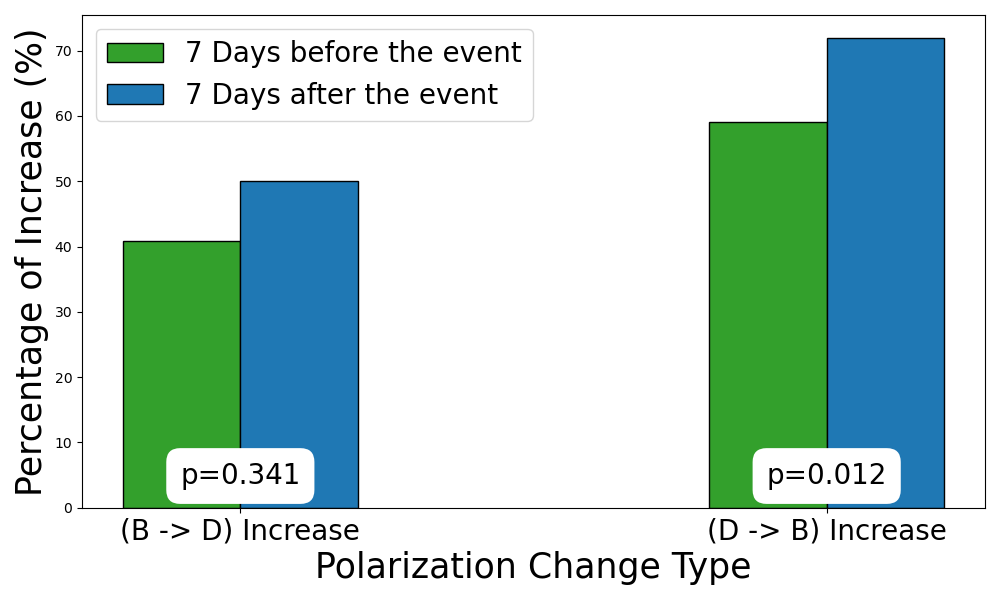}
        \caption{Hurricane Ida}
        \label{fig:plot4}
    \end{subfigure}
    \hfill
    \begin{subfigure}[b]{0.32\textwidth}
        \centering
        \includegraphics[width=\linewidth]{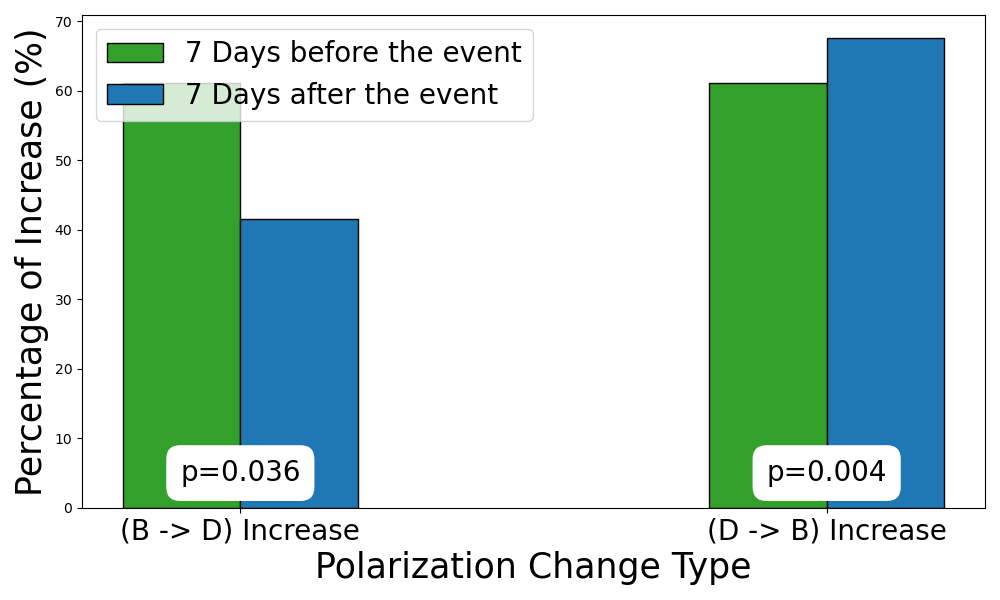}
        \caption{Heat Wave}
        \label{fig:plot5}
    \end{subfigure}
    \hfill
    \begin{subfigure}[b]{0.32\textwidth}
        \centering
        \includegraphics[width=\linewidth]{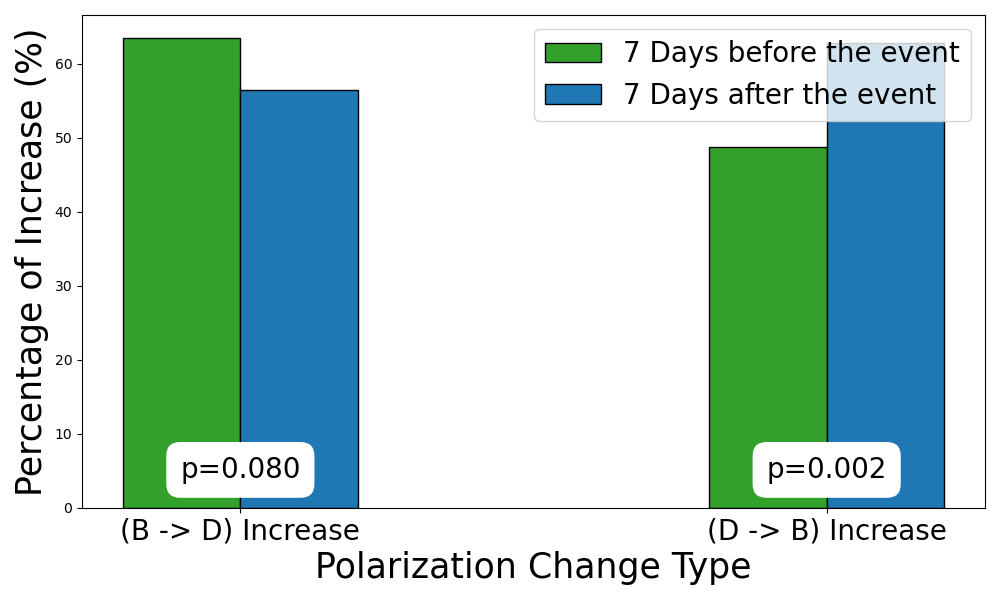}
        \caption{IPCC Annual Report}
        \label{fig:plot6}
    \end{subfigure}

    \caption{Temporal dynamics of polarization scores in response to events. The top trio of subfigures (a-c) illustrates changes in polarization preceding and following events related to gun control, with '(P $\rightarrow$ A)' denoting shifts from pro-to-anti gun control stance and '(A $\rightarrow$ P)' denoting shifts from anti-to-pro stance. The bottom trio (d-f) depicts similar changes for climate change-related events, with  '(B $\rightarrow$ D)' representing shifts from believe-to-disbelieve in climate change and '(D $\rightarrow$ B)' for shifts from disbelieve-to-believe stance direction. P-values indicate the statistical significance of changes, confirming that these shifts are not random but are influenced by the events.}
    
    % \caption{Polarization score changes before and after an event has occurred. The top three figures represent events related to gun control topic and the bottom three events represent those related to climate change.}
    \label{fig:polarization_timeline}
\end{figure*}

Our analysis reveals a temporal sensitivity in polarization, underscored by fluctuations that correspond closely with real-world events and influencer-led conversations. Specifically, our temporal examination—illustrated in Figures~\ref{fig:conversation_distribution} and~\ref{fig:polarization_timeline}—highlights notable shifts in polarization within the contexts of gun control and climate change discourse. 

For instance, in Figure~\ref{fig:gun_tweets: conv} during the summer of 2022, the gun control dataset exhibited significant spikes in the daily Twitter conversation, correlating with high-profile shooting incidents in the United States. This period, notably marked by the tragic Texas Robb Elementary School shooting, saw intensified influencer activity leading conversations that either amplified or attempted to bridge divides in public sentiment. Similarly, in the climate change discourse (Figure~\ref{fig:climate_tweets: conv}), pivotal events such as the release of the Intergovernmental Panel on Climate Change (IPCC) report and the United Nations Climate Change Conference were mirrored by peaks, pointing to the reactive nature of online discourse to global climate events. 

% Following these events, Figure~\ref{fig:polarization_timeline} illustrates how influential-led conversation changes the dynamic of polarization. For both, gun-control and climate-change datasets, we observe that when a certain event occurs, the polarization score increases compared to the situation it was before the event happened. These findings not only demonstrate the temporal alignment of polarization with external events but also highlight the role of influencers in steering the conversation, thereby affecting the polarization landscape.

Following these events, Figure~\ref{fig:polarization_timeline} illustrates how influential-led conversation changes the dynamic of polarization. For both, gun-control and climate-change datasets, we observe that when a certain event occurs, the polarization score increases compared to the situation it was before the event happened. Statistical significance tests confirm that many of these observed shifts are significant, reinforcing polarization's temporal alignment with external events. These observations highlight the critical role of influencers in steering the conversation and the powerful impact that influencers and real-world events can have affecting the polarization landscape.

%%%% CLIMATE TABLE %%%%%

\begin{table*}[h!]
\small
\centering
\caption{Climate Change Dataset: The effect of a single conversation by influential users on daily polarization score.}
\begin{adjustbox}{width=\textwidth}
% \usepackage{multirow}

\begin{tabular}{|l|l|r|r|r|r|r|r|}
\hline
\textbf{User} & \begin{tabular}[c]{@{}l@{}}\textbf{User's} \\ \textbf{Stance}\end{tabular} & \multicolumn{1}{l|}{\begin{tabular}[c]{@{}r@{}}\textbf{Number of Tweets}\\ \textbf{in the Conversation}\end{tabular}} & \multicolumn{1}{l|}{\begin{tabular}[c]{@{}r@{}}\textbf{Number of Followers}\\ \textbf{in the Conversation}\\ \textbf{(\#disbelieve / \#believe)}\end{tabular}} & \multicolumn{1}{l|}{\begin{tabular}[c]{@{}r@{}}\textbf{Number of Followers}\\ \textbf{Interacting on That Day}\\ \textbf{(\#disbelieve / \#believe)}\end{tabular}} & \multicolumn{1}{l|}{\begin{tabular}[c]{@{}l@{}}\textbf{Stance}\\ \textbf{Direction}\end{tabular}} & \multicolumn{1}{l|}{\begin{tabular}[c]{@{}r@{}}\textbf{Polar. Score}\\ \textbf{Without Conv.}\end{tabular}} & \multicolumn{1}{l|}{\begin{tabular}[c]{@{}r@{}}\textbf{Polar. Score}\\ \textbf{With Conv.}\end{tabular}} \\ \hline
\multirow{2}{*}{User 1} & \multirow{2}{*}{believe} & \multirow{2}{*}{104} & \multirow{2}{*}{41 / 50} & \multirow{2}{*}{904 / 1,332} & disbelieve $\rightarrow$ believe & {-0.091} & {-0.135} \\ \cline{6-8} 
 &  &  &  &  & believe $\rightarrow$ disbelieve & 0.463 & 0.398 \\ \hline
\multirow{2}{*}{User 2} & \multirow{2}{*}{believe} & \multirow{2}{*}{187} & \multirow{2}{*}{131 / 24} & \multirow{2}{*}{457 / 333} & disbelieve $\rightarrow$ believe & {0.115} & {0.185} \\ \cline{6-8} 
 &  &  &  &  & believe $\rightarrow$ disbelieve & 0.378 & 0.402 \\ \hline
\multirow{2}{*}{User 3} & \multirow{2}{*}{disbelieve} & \multirow{2}{*}{157} & \multirow{2}{*}{68 / 14} & \multirow{2}{*}{302 / 100} & disbelieve $\rightarrow$ believe & {-0.137} & {-0.136} \\ \cline{6-8} 
 &  &  &  &  & believe $\rightarrow$ disbelieve & {0.633} & {0.471} \\ \hline
\multirow{2}{*}{User 4} & \multirow{2}{*}{believe} & \multirow{2}{*}{134} & \multirow{2}{*}{51 / 37} & \multirow{2}{*}{356 / 356} & disbelieve $\rightarrow$ believe & {0.149} & {0.181} \\ \cline{6-8} 
 &  &  &  &  & believe $\rightarrow$ disbelieve & {-0.228} & {-0.247} \\ \hline
\multirow{2}{*}{User 5} & \multirow{2}{*}{disbelieve} & \multirow{2}{*}{66} & \multirow{2}{*}{43 / 2} & \multirow{2}{*}{395 / 86} & disbelieve $\rightarrow$ believe & {0.144} & {0.152} \\ \cline{6-8} 
 &  &  &  &  & believe $\rightarrow$ disbelieve & {0.248} & {0.237} \\ \hline
\end{tabular}

\end{adjustbox}
\label{table-climate-conv-details}
\end{table*}

%%%% GUN CONTROL TABLE %%%%%

\begin{table*}[h!]
\small
\centering
\caption{Gun Control: The effect of a single conversation by influential users on daily polarization score.}
\begin{adjustbox}{width=\textwidth}
% \usepackage{multirow}

\begin{tabular}{|l|l|r|r|r|r|r|r|}
\hline
\textbf{User} & \begin{tabular}[c]{@{}l@{}}\textbf{User's} \\ \textbf{Stance}\end{tabular} & \multicolumn{1}{l|}{\begin{tabular}[c]{@{}r@{}}\textbf{Number of Tweets}\\ \textbf{in the Conversation}\end{tabular}} & \multicolumn{1}{l|}{\begin{tabular}[c]{@{}r@{}}\textbf{Number of Followers}\\ \textbf{in the Conversation}\\ \textbf{(\#anti / \#pro)}\end{tabular}} & \multicolumn{1}{l|}{\begin{tabular}[c]{@{}r@{}}\textbf{Number of Followers}\\ \textbf{Interacting on That Day}\\ \textbf{(\#anti / \#pro)}\end{tabular}} & \multicolumn{1}{l|}{\begin{tabular}[c]{@{}l@{}}\textbf{Stance}\\ \textbf{Direction}\end{tabular}} & \multicolumn{1}{l|}{\begin{tabular}[c]{@{}r@{}}\textbf{Polar. Score}\\ \textbf{Without Conv.}\end{tabular}} & \multicolumn{1}{l|}{\begin{tabular}[c]{@{}r@{}}\textbf{Polar. Score}\\ \textbf{With Conv.}\end{tabular}} \\ \hline
\multirow{2}{*}{User 1} & \multirow{2}{*}{pro} & \multirow{2}{*}{746} & \multirow{2}{*}{334 / 325} & \multirow{2}{*}{1,211 / 1,040} & anti $\rightarrow$ pro & {0.147} & {0.252} \\ \cline{6-8} 
 &  &  &  &  & pro $\rightarrow$ anti & {0.247} & {-0.200} \\ \hline
\multirow{2}{*}{User 2} & \multirow{2}{*}{anti} & \multirow{2}{*}{1,333} & \multirow{2}{*}{497 / 391} & \multirow{2}{*}{856 / 707} & anti $\rightarrow$ pro & {0.079} & {0.229} \\ \cline{6-8} 
 &  &  &  &  & pro $\rightarrow$ anti & {0.220} & {-0.087} \\ \hline
\multirow{2}{*}{User 3} & \multirow{2}{*}{pro} & \multirow{2}{*}{57} & \multirow{2}{*}{20 / 22} & \multirow{2}{*}{312 / 356} & anti $\rightarrow$ pro & 0.085 & 0.099 \\ \cline{6-8} 
 &  &  &  &  & pro $\rightarrow$ anti & {-0.022} & {0.121} \\ \hline
\multirow{2}{*}{User 4} & \multirow{2}{*}{anti} & \multirow{2}{*}{164} & \multirow{2}{*}{37 / 100} & \multirow{2}{*}{176 / 286} & anti $\rightarrow$ pro & {0.247} & {0.121} \\ \cline{6-8} 
 &  &  &  &  & pro $\rightarrow$ anti & {0.204} & {0.241} \\ \hline
\multirow{2}{*}{User 5} & \multirow{2}{*}{pro} & \multirow{2}{*}{129} & \multirow{2}{*}{16 / 50} & \multirow{2}{*}{867 / 1,879} & anti $\rightarrow$ pro & {0.043} & {0.041} \\ \cline{6-8} 
 &  &  &  &  & pro $\rightarrow$ anti & {0.069} & {0.085} \\ \hline
\end{tabular}

\end{adjustbox}
\label{table-gun-conv-details}
\end{table*}

%% GUN CONTROL

\begin{table*}[h!]
\small
\centering
\caption{Gun Control Dataset: Impact of conversation initialized by influential authors (Twitter users) on polarization scores. The '\% increase' or '\% decrease' denotes the percentage of cases, (how) polarization score changes upon adding a conversation compared to the total number of conversations.}
\begin{adjustbox}{width=\textwidth}

\begin{tabular}{|l|l|l|r|r|r|}
\hline
\textbf{Twitter User} & \textbf{User Type} & \begin{tabular}[c]{@{}l@{}}\textbf{Stance}\\ \textbf{(Gun Control)}\end{tabular} & \begin{tabular}[c]{@{}r@{}}\textbf{Number of} \\ \textbf{Conversations}\end{tabular} & \begin{tabular}[c]{@{}r@{}}\textbf{Polarization: Change}\\ \textbf{(pro $\rightarrow$ anti)}\end{tabular} & \begin{tabular}[c]{@{}r@{}}\textbf{Polarization: Change}\\ \textbf{(anti $\rightarrow$ pro)}\end{tabular} \\ \hline
User 1 & Media Outlet & anti & 5 & 100.0 \% - decrease & 100.0 \% - increase \\ \hline
User 2 & Media Outlet & pro & 6 & 66.67 \% - decrease & 100.0 \% - decrease \\ \hline
User 3 & Commentator & pro & 9 & 88.89 \% - decrease & 77.78\% - decrease \\ \hline
User 4 & Journalist & anti & 18 & 88.89 \% - decrease & 61.11\% - increase \\ \hline
User 5 & Politician & pro & 7 & 85.71 \% - increase & 85.71\% - decrease \\ \hline
\end{tabular}

\end{adjustbox}
\label{table-gun}
\end{table*}

%% CLIMATE CHANGE

\begin{table*}[h!]
\small
\centering

\caption{Climate Change Dataset: Impact of conversation initialized by influential authors (Twitter users) on polarization scores. The '\% increase' or '\% decrease' denotes the percentage of cases, (how) polarization score changes upon adding a conversation compared to the total number of conversations.}
\begin{adjustbox}{width=\textwidth}
% \usepackage{multirow}

\begin{tabular}{|l|l|l|r|r|r|}
\hline
\textbf{Twitter User} & \textbf{User Type} & \begin{tabular}[c]{@{}l@{}}\textbf{Stance}\\ \textbf{(Climate Change)}\end{tabular} & \begin{tabular}[c]{@{}r@{}}\textbf{Number of} \\ \textbf{Conversations}\end{tabular} & \begin{tabular}[c]{@{}r@{}}\textbf{Polarization: Change}\\ \textbf{(believe $\rightarrow$ disbelieve)}\end{tabular} & \begin{tabular}[c]{@{}r@{}}\textbf{Polarization: Change}\\ \textbf{(disbelieve $\rightarrow$ believe)}\end{tabular} \\ \hline
User 1 & Scientific Org. & believe & 5 & 60.00\% - decrease & 100.0 \% - increase \\ \hline
User 2 & Public Figure & disbelieve & 5 & 80.00\% - decrease & 100.0 \% - increase \\ \hline
User 3 & Media Personality & disbelieve & 8 & 50.00\% - increase & 87.5 \% - decrease \\ \hline
User 4 & Politician & believe & 7 & 85.71\% - decrease & 57.14 \% - decrease \\ \hline
User 5 & Media Outlet & believe & 41 & 51.22\% - decrease & 85.37 \% - increase \\ \hline

\end{tabular}

\end{adjustbox}
\label{table-climate}
\end{table*}

\begin{table*}[h!]
\small
\centering
\caption{Details of conversations' effects based on stances for both gun control and climate change. The '\% increase' or '\% decrease' denotes the percentage of cases, (how) polarization score changes upon adding a conversation compared to the total number of conversations.}
\begin{adjustbox}{width=1.55\columnwidth}

\begin{tabular}{|c|lr|rr|}
\hline
\textbf{Dataset} & \multicolumn{1}{l|}{\textbf{Stance}} & \multicolumn{1}{l|}{\textbf{\begin{tabular}[c]{@{}r@{}}Number of\\ Conversations\end{tabular}}} & \multicolumn{2}{c|}{\textbf{Polarization Score: Change}} \\ \hline
\multirow{3}{*}{Gun Control} & \multicolumn{2}{l|}{} & \multicolumn{1}{c|}{\textbf{(pro $\rightarrow$ anti)}} & \multicolumn{1}{c|}{\textbf{(anti $\rightarrow$ pro)}} \\ \cline{2-5} 
 & \multicolumn{1}{l|}{pro} & 1,940 & \multicolumn{1}{r|}{63.40\% - decrease} & 52.99\% - increase \\ \cline{2-5} 
 & \multicolumn{1}{l|}{anti} & 1,704 & \multicolumn{1}{r|}{55.52\% - decrease} & 53.28\% - decrease \\ \hline
\multirow{3}{*}{Climate Change} & \multicolumn{2}{l|}{} & \multicolumn{1}{c|}{\textbf{(believe $\rightarrow$ disbelieve)}} & \multicolumn{1}{c|}{\textbf{(disbelieve $\rightarrow$ believe)}} \\ \cline{2-5} 
 & \multicolumn{1}{l|}{believe} & 1,561 & \multicolumn{1}{r|}{50.99\% - decrease} & 64.25\% - increase \\ \cline{2-5} 
 & \multicolumn{1}{l|}{disbelieve} & 596 & \multicolumn{1}{r|}{50.67\% - decrease} & 58.22\% - decrease \\ \hline
\end{tabular}

\end{adjustbox}
\label{table-stance-level-details}
\end{table*}

\subsection{Effect of an Influential-led Conversation} 

The shifts in polarization are quantified by examining the changes in polarization scores when specific influencer-led conversation is either included or excluded from the analysis. This approach helps isolate the direct impact of a single conversation on the overall polarization score that day.

For example, as illustrated in Table~\ref{table-climate-conv-details}, the removal of a conversation by 'User 3', a prominent disbeliever in climate change, resulted in a notable shift in the polarization scores. Specifically, the polarization score for the believer-to-disbeliever direction increased from 0.471 to 0.633, indicating a more pronounced divide when the disbeliever's influence was absent from the conversation. This change, representing a magnitude of 0.162 in the polarization score, underscores the significant role that influencer-led interactions play in shaping public sentiment and polarization.

Similarly, in the context of gun control as detailed in Table~\ref{table-gun-conv-details}, the exclusion of a conversation led by an anti-gun control user (i.e. 'User 2') resulted in a decrease in the anti-pro polarization score. This demonstrates how conversations led by users opposed to gun control amplify polarization towards anti-gun stances when present, and reduce it when removed.

\subsection{Insight of Influential Users}

In this section, we examine the collective influence of conversations initiated by influential users on shaping public sentiment and polarization across various themes. We employ a systematic approach to quantify the impact of these users by assessing the frequency and effect of their conversations on polarization shifts.

Tables~\ref{table-gun} and ~\ref{table-climate} present data on influential users who actively engage in discussions and initiate several conversations related to gun control and climate change, respectively. These tables reveal the percentage of conversations that increase or decrease polarization, given a specified number of conversations within the context of these topics, thereby illuminating the effects of influential users on social media dynamics.

% In this section, we focus on the collective impact of conversations initiated by influential users, assessing how these conversations shape public sentiment and polarization across different themes. We systematically quantify the influence of these users by evaluating the frequency and impact of their conversations on polarization changes.

% These tables reveal that influencers with stances opposing gun control or skeptical of climate change consistently drive polarization in both directions in almost all cases. This pattern underscores the power of their narratives to fortify opposition and dilute support for these causes. In contrast, influencers who advocate for gun control or proactive measures against climate change generally promote a reduction in polarization, encouraging a discourse that both attracts supporters and moderates public opinion.

\subsection{Insight of Conversations Within Stance-Group}
Table~\ref{table-stance-level-details} provides a more comprehensive view by analyzing all the conversations initiated by users within a certain stance. Interestingly, we observe that conversations led by users with anti or disbeliever stances often result in decreased polarization compared to their opposing stance group.  However, it's important to note the observational nature of our approach; while it reveals significant insights into the dynamics of influencer conversations, it does not assert direct causation. Overall, the insights gleaned from these influencer-led conversations across different thematic areas demonstrate the critical role of influential users in modulating the affective landscape of online communities. 

% This suggests that when aggregating all conversational instances, the anticipated increase in polarization due to influential anti or disbeliever discourse does not consistently materialize.

%By identifying the conditions under which influencer-led conversations amplify or temper polarization, our study contributes to the development of more harmonious digital discourse environments.

\section{Conclusion}

This study provides a detailed analysis of the mechanics of affective polarization on social media, particularly highlighting the crucial role of influential users in modulating public sentiment on platforms like Twitter. By implementing a counterfactual framework, which evaluates scenarios with and without specific influencer-led conversations, we offer a unique methodological contribution that allows us to isolate and quantify the influence of such conversations on the dynamics of polarization. Our findings demonstrate that influencers have the potential to amplify or mitigate divisiveness across polarizing topics like gun control and climate change. We reveal how subtle shifts in influencer-driven dialogues can significantly affect public discourse by employing a comparative analysis coupled with computational techniques, such as subgraph construction. This approach deepens our understanding of how specific conversations impact polarization that are critical to social media dynamics. 
%Future research should delve deeper into identifying the attributes of influencers—such as their political ideology, beliefs, demographic factors, and the nature of their discussions—that drive their impact on polarization. This line of inquiry could provide a more nuanced understanding of how different types of influencers shape public sentiment.

Despite these promising results, our study faces methodological challenges, particularly due to limitations inherent in calculating the polarization scores based on sentiment analysis. This depends on the presence of positive or negative words in tweets and it is negatively impacted by the sporadic nature of meaningful social media interactions. 
%The small scale of interactions within specific conversations compared to the vast daily interactions across the platform posed significant challenges in achieving statistical significance. 
These challenges underscore the complexity of capturing the sophisticated landscape of online discourse and point to the need for developing more refined metrics and methodologies that leverage current advancements in language understanding that go beyond sentiment analysis. 
%Moreover, this research advances scholarly understanding of online polarization by pioneering the use of counterfactual scenarios, significantly illuminating the complex dynamics of social media influence.

% ACKNOWLEDGEMENT % --------------------------------------------------------
\section*{Acknowledgement}
{The research was sponsored by the Army Research Office and was accomplished under Grant Number W911NF-22-1-0035. The views and conclusions contained in this document are those of the authors and should not be interpreted as representing the official policies, either expressed or implied, of the Army Research Office or the U.S. Government. The U.S. Government is authorized to reproduce and distribute reprints for Government purposes notwithstanding any copyright notation herein.}

\bibliographystyle{ieeetr}
\bibliography{references}

\end{document}